# What do Transgender Software Professionals say about a Career in the Software Industry?

**Ronnie de Souza Santos[1,2], Brody Stuart-Verner[1], Cleyton Magalhães[2]**
[1] Cape Breton University – Canada
[2] CESAR School – Brazil
Email: ronnie_desouza@cbu.ca, cvcm@cesar.school, brody_verner@cbu.ca

**ABSTRACT**. Diversity is an essential aspect of software development because technology influences almost every aspect of modern society, and if the software industry lacks diversity, software products might unintentionally constrain groups of individuals instead of promoting an equalitarian experience to all. In this study, we investigate the perspectives of transgender software professionals about a career in software engineering as one of the aspects of diversity in the software industry. Our findings demonstrate that, on the one hand, trans people choose careers in software engineering for two primary reasons: a) even though software development environments are not exempt from discrimination, the software industry is safer than other industries for transgenders; b) trans people occasionally have to deal with gender dysphoria, anxiety, and fear of judgment, and the work flexibility offered by software companies allow them to cope with these issues more efficiently.

## I. INTRODUCTION

Technology plays an important role in people's lives, influencing almost every aspect of our society, such as work, education, politics, and leisure. As our society is diverse and plural, software engineering is expected to reflect this diversity to ensure that software products provide equal support to people rather than unintentionally constrain some groups of individuals [1]. However, the reality is that the software industry is facing a *diversity crisis* [1][2], which means that software development environments are highly homogeneous, i.e., the opposite of the society it serves. In this sense, the software industry needs a workforce concerned with improving equity, diversity, and inclusion in software engineering careers to enable software professionals to have a fair, equalitarian, and just experience, allowing them to improve equity, diversity, and inclusion aspects in the software products they implement [1][3].

Recent studies demonstrate that the post-pandemic work in software companies, which is based on remote and hybrid structures, has benefited several software professionals from underrepresented groups who previously faced various challenges to work in software engineering careers, such as caregivers (especially mothers) [4][5], people with disabilities [4], and LGBTQIA+ individuals (especially transgender people) [6]. In particular, for transgender software professionals, remote work is desirable because: a) transgender people are historically one of the most discriminated professionals in hiring processes and work environments; b) remote environments benefit them with identity disclosure and the autonomy to disengage and re-engage with their peers, i.e., having more control of their interactions [7].

In this study, we investigate the perspectives of transgender software professionals about a career in software engineering, including their experience working with software development in the post-pandemic period. We interviewed three trans software professionals about their work in the software industry, the obstacles they face in their careers, and how they are affected by remote work structures. Our discussions aim to promote awareness about diversity in software engineering careers and support software companies in improving inclusion in software development environments based on the premise that more inclusive software environments are essential to the development of more inclusive software [6].

## II. SOFTWARE ENGINEERING *IS A* HETEROSEXUAL *MAN'S MAN'S MAN'S WORLD*

Programming, one of the core activities of software development, was initiated by a woman, and this work remained to be performed by women for decades [1]. However, this is not the case today, as men are reported to represent over 75% of the workforce in software engineering [8]. For other underrepresented groups, e.g., the LGBTQIA+ community, the number of individuals working in the software industry is very small compared to the number of cisgender heterosexuals, which makes software engineering a cisgender *heterosexual man's world* [1][6][8].

The lack of diversity in software engineering has produced several consequences over the years. Computational algorithms and decision-making systems (e.g., machine learning-based algorithms) have been reported to be systematically discriminatory against non-white communities as biased implementations prioritize the needs of white individuals while excluding other ethnic groups from services and opportunities [3]. Gender imbalances in software engineering foster hostile and sexist environments that hinder opportunities for non-male individuals to succeed in this field [8]. The diversity problem in software engineering is even perceived at the educational level since female and LGBTQIA+ students are more likely to drop software engineering and co-related courses due to a low sense of belonging and outright bullying [1].

In general, pursuing a career in software engineering comes with several challenges that professionals must overcome, including dealing with constant technological advances (e.g., experience, qualification), adapting to team dynamics (e.g., types of leadership, communication faults, collaboration issues), and dealing with conflicts with clients or products (e.g., unclear requirements, feature prioritization) [9]. However, software professionals from underrepresented groups (e.g., those who identify as women, non-binary, and transgender) face even more difficulties in the software industry as the unwelcome environment makes them frequently question their abilities to work in this area [1]. In this scenario, transgender software professionals have additional challenges to overcome as they struggle with how colleagues treat and respect their identities [7] [10]. In summary, the software industry is primarily known as a *man's world*, as individuals from underrepresented groups (e.g., women and LGBTQIA+) face disadvantages that are rarely experienced by cisgender heterosexual men [1][3][7][8].

## III. METHOD

We conducted a qualitative study based on interviews to explore the experiences of transgender software professionals and understand their perspectives on a career in the software industry. Based on the qualitative nature of our study, we are not focusing on any statistical generalization of results; instead, we claim that practitioners can use the findings obtained from our qualitative analysis to improve their knowledge about diversity, allowing them to compare the experiences reported in this study with the reality of their work environments.

### A. SELECTING PARTICIPANTS

We followed the recommendations for treating software professionals as a hidden population [11]. Transgender software professionals are considered a hidden population because a sample of individuals from this population cannot be easily defined or enumerated based on existing knowledge [11]. Moreover, individuals from the LGBTQIA+ communities are commonly treated as a hidden population because many of their members are not comfortable with discussing aspects of their sexuality due to the risk of being exposed to structural and social discrimination [12]. Therefore, we applied two techniques to identify and select participants for this study. First, we applied convenience sampling [11] to select participants based on their availability, using our extensive network of researchers and practitioners to identify transgender software professionals interested in participating in our study. Second, we used snowballing sampling [11] by asking participants who were interviewed in this study to recommend other professionals that could also participate. Using these two sampling techniques, we identified seven transgender software professionals. However, only three of them accepted to be interviewed about their careers.

### B. DATA COLLECTION

We used unstructured interviews as our data collection strategy. Unstructured interviews [13] are particularly useful for obtaining relevant data from participants' stories and experiences in a flexible yet reliable way. An unstructured interview is initiated with a group of general questions about a theme, and subsequently, it follows a list of topics to be discussed by the interviewee instead of focusing on a list of structured questions, which is common in semi-structured interviews. In this study, we started by asking participants to comment on their experience working in software teams (broad open question); then, we continued discussing aspects of their careers based on the following aspects:

    a) Their reason for opting to work with software development.
    b) The benefits and challenges of working in software companies.
    c) The benefits and the challenges of working with software teams.
    d) Their expectations for the future in the software industry.

### C. DATA ANALYSIS

We used thematic analysis to analyze the data collected in the interviews. Thematic analysis is commonly used to identify, analyze, and interpret interview quotations and reveal key themes within qualitative data [14]. Similar to our data collection strategy, we selected thematic analysis in this study based on its flexibility, which allows the analysis of small datasets (e.g., 1-2 participants) or a large number of interviews [14]. Using thematic analysis, we broke down quotes from the interviews into codes and posteriorly grouped them into themes or categories. Following this, we identified connections among these themes in the same interview and across interviews. This process resulted in a description of the perspectives of transgender software professionals and their experiences in the software industry.

*D. ETHICS*

The personal information of participants (e.g., name, contact, or employer) was de-identified to guarantee anonymity. In addition, participants were informed about using the collected data for scientific purposes and agreed to participate in the interviews.

## IV. FINDINGS

We interviewed three software professionals who identify as transgender. Table 1 summarizes our sample. Following the thematic analysis, we present our findings based on the narrative of three career aspects of transgender software professionals:
a) Choosing a career in the software industry.
b) Interacting with co-workers in the software team.
c) Impacts of their experience in the software products.

**Table 1. Demographics**

| Participant | Gender Identity | Sexual Orientation | Role | Experience |
|---|---|---|---|---|
| P01 | Transgender man | Asexual | Software Designer | 3 years |
| P02 | Transgender woman | Pansexual | Programmer | 1 year |
| P03 | Transgender woman | Heterosexual | Software Tester/QA | 1 year |

*A. TRANSGENDER SOFTWARE PROFESSIONALS FEEL SAFER WORKING IN THE SOFTWARE INDUSTRY*

Transgender people still face severe discrimination and violence in our society, including in their workplaces. However, pursuing a career in software development is becoming an alternative for them to work in a safer space. The reason why trans people choose a career in software engineering is related to the prospect that most of the work in software development can be done remotely, which allows them to avoid several discriminatory situations. P01 discussed that working from home helps trans people to be safer as they can avoid discrimination resulting from commuting between home and work. P02 reported that she abandoned a career in health to pursue a career in software engineering because working with technology allows her to choose the best day and time to go to the office, e.g., avoiding rush hours.

In addition, being a transgender person means having to deal with several personal issues, including gender dysphoria, anxiety, and fear of judgment. Currently, while the software industry is experiencing an increase in the number of remote and hybrid positions, trans people are not only focused on a career in software engineering but also opting for companies that provide them with high levels of work freedom and flexibility. Working remotely allows trans software professionals to have better control of their identities by deciding when and for how long they will use their cameras or if they will simply interact using other technologies. Moreover, simple yet empowering actions are better managed in virtual environments, such as stating the pronouns to which they identify or how they want to be called. P02 discussed how remote structures support transgender software professionals by allowing them to gradually be comfortable around their teammates until the point they feel accepted and welcome to interact more directly.

> "I was working in Health, then I thought with myself, I should go to Technology because it will be safer for me (due to remote work)." (P03)

> "Working from home is viable, it provides you with more safety, makes you feel more confident." (P01)

> "Because sometimes I get dysphoric and then I have to set the camera in angle that my body will not be completely visible." (P01)

> "I started working remotely and then when I had to go to the office, I felt more comfortable because my colleagues already knew me." (P02)

### B. TRANSGENDER SOFTWARE PROFESSIONALS FEEL UNWELCOMED IN THE SOFTWARE INDUSTRY

Although working in the software industry provides transgender software professionals with benefits related to safety, flexibility, and more control of their identity, they face challenges in the workspace that other software professionals do not experience. Most of these challenges relate to teamwork aspects, including involvement with colleagues, team engagement, and sense of belonging. P01 discusses that trans software professionals are apprehensive about being unwelcomed when they join software teams, and they fear that teams that are not diverse will be inconsiderate of their ideas and the outcomes of their work. P03 discusses her fear of being put in an uncomfortable situation when interacting with teams that are mainly composed of male individuals. P01 reported difficulty in getting his colleagues' attention several times, even when he is the most experienced person about a feature under development.

Further, transgender software professionals emphasize that organizations should lead discussions on the importance of diversity for software development. They also highlight the role of software companies in promoting inclusiveness and protecting people from underrepresented groups from unpleasant circumstances. P01 reports that software companies play an essential role in shaping diversity awareness; therefore, this responsibility should not be assigned only to transgender (and LGBTQIA+) individuals but to everyone in the company, as diversity positively impacts the software products released to clients and users. P03 reinforces the importance of managers and leaders informing the team on how to respect and work harmoniously with individuals from underrepresented groups, in particular, transgender people.

> "I was really stressed about how I would be accepted when I joined my team which is mostly composed by cisgender straight men." (P01)

> "At first it was hard, I was really afraid not being accepted, of having someone being unpleasant to me." (P03)

> "I criticized the committee for always waiting for LGBTQIA+ people to raise discussions on diversity in the company." (P01)

> "There is more violence in the workspace than just mistaking our pronouns, having your effort unacknowledged for not having a heteronormative appearance is one of them." (P01)

> "It's literally preparing the team to receive the new member and explain that as a trans person I use these pronouns." (P03)

### C. TRANSGENDER SOFTWARE PROFESSIONALS INCREASE CREATIVITY AND PROBLEM-SOLVING IN SOFTWARE DEVELOPMENT

Our findings demonstrate that transgender software professionals play an important role in software development, as these individuals contribute to the promotion of diversity and inclusion aspects in the software development, thus, helping the software team to reproduce these aspects into software products, including but not restricted to:

a) Providing analytical perspective on UX/UI and how the software communicates with diverse users.
b) Improving requirements based on a variety of experiences.
c) Helping to identify and decrease biases in the software development process and in the software outcomes.
d) Contributing to the team's ability to deal with complex problems.

P01 discuss that when transgender people pursue a career in software engineering, they tend to incorporate their own world experience to make software products and processes less aggressive and more empathetic. P02 reinforces the importance of diversity in software teams to create solutions for complex problems.

> "We think about how to make the software less aggressive or more inclusive because we bring more social criticism the to process." (P01)

> "Exploring is a big deal in software development, so it is important to have a variety of perspectives." (P02)

> "Diversity is very important in a group that is trying to come up with a solution." (P02)

## V. DISCUSSIONS

### A. COMPARING FINDINGS WITH THE LITERATURE

The literature about transgender people working in the software industry is scarce. In general, we confirmed the findings presented in [7] by demonstrating that remote work is more inclusive for transgender software professionals. In addition, we extend these findings and demonstrate that the interest of trans people in software engineering careers is increasing, mainly because software companies are more likely to implement remote and hybrid work structures that can benefit these individuals significantly. We also confirmed what is discussed in [1] since our findings highlighted that, similar to individuals from other underrepresented groups, transgender software professionals also fear exclusion and rejection in software development environments. Transgender software professionals worry about having their abilities questioned by the team based on prejudgments related to their gender expression or sexual orientation. However, the possibility of having more control over identity in virtual environments helps these professionals to deal with such feelings as remote structures provide them with the flexibility of choosing how and when to interact with their peers (e.g., text, video, or in-person). Finally, as far as we know, this is the first study to reveal that physical well-being (e.g., feeling safer) is one of the main reasons why transgender people are pursuing careers in software engineering.

### B. IMPLICATIONS TO ACADEMIA AND PRACTICE

Regarding implications for academia, we start by highlighting the lack of studies that address the community of transgender software professionals and their experiences working in software development. Although the number of studies investigating aspects of diversity and inclusion in software engineering seems to be gradually increasing [2], transgender people are rarely included in software engineering studies [7], especially due to the difficulty of sampling this population (i.e., hidden population). We emphasize that researchers must develop strategies to reach these professionals to investigate the challenges they face in software development environments, including those already mentioned in the present study, e.g., teamwork challenges. In addition, we highlight the importance of investigating the contributions of these professionals to promote diversity and inclusion in software development and software products, including aspects of creativity and problem-solving.

As for practitioners, we highlight the importance of diversity in software teams to promote innovation [6]. Innovation is the key to successful growth and competitive advantage in the software industry [15], and our findings revealed that transgender software professionals could provide significant support in creative processes and the implementation of solutions for complex problems. More importantly, these professionals are essential in developing systems that do not discriminate against groups of individuals in our diverse society. Therefore, we recommend that software companies develop strategies to improve the attractiveness of software engineering careers to professionals from underrepresented groups, for instance, by increasing work flexibility (e.g., improving remote and hybrid work structures), promoting diversity awareness, and stimulating inclusiveness in the software development environments.

## VI. CONCLUSIONS

The present study focused on discussing the perspective of transgender software professionals about a career in software engineering. Our results demonstrated that the interest of trans people in pursuing a career in software engineering is increasing because of the flexibility resulting from remote and hybrid work offered by software companies that allows them to feel safer and manage social interactions according to their limitations. However, these professionals face challenges working with homogeneous teams, especially regarding the difficulty of interacting with co-workers. The software industry is expected to improve its practices and promote more diversity and inclusion to produce software that serves our plural society and does not discriminate against groups of individuals. Transgender software professionals contribute significantly to this goal since their experience supports the development of more inclusive software products.

## AUTHORS

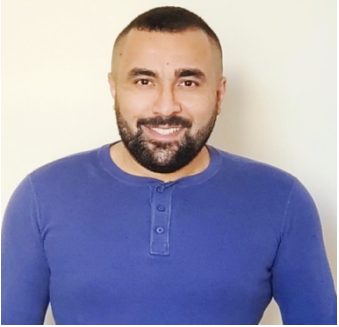

**Ronnie de Souza Santos** is an assistant professor at Cape Breton University in Canada. He received his Ph.D. in computer science from the Federal University of Pernambuco (Brazil) and completed 2 years of postdoctoral fellowship in software engineering at Dalhousie University in Canada. His research interests are equity, diversity, and inclusion in software engineering and software development practices. He is an adjunct professor at CESAR School (Brazil).

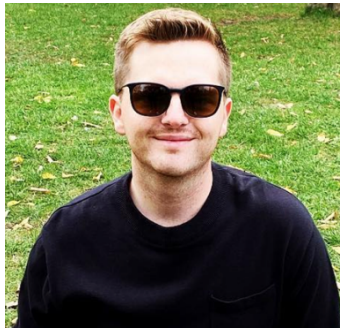

**Brody Stuart-Verner** is a writer, podcaster and LGBTQIA+ activist. He graduated from Mount Saint Vincent University in 2018, earning a Bachelor of Arts in Communication Studies. He is currently employed in an administrative capacity at Cape Breton University in Canada. His research interests include political communication and crisis management.

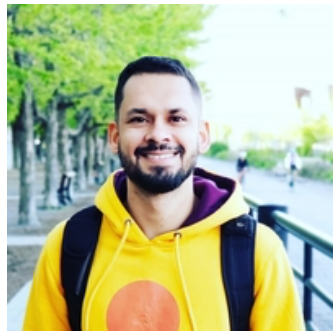

**Cleyton Magalhes** is a professor in the Post Baccalaureate program in Agile Testing at CESAR School in Brazil. He received his Ph.D. in computer science from the Federal University of Pernambuco (Brazil). He is a software QA at the Recife Center for Advanced Studies and Systems (CESAR). His research interests include human aspects of software engineering and software testing.